\documentclass[prb,twocolumn,showpacs]{revtex4}

\def\betasf{$\beta''$-(BEDT-TTF)$_2$SF$_5$CH$_2$CF$_2$SO$_3$}
\def\cuscn{$\kappa$-(BEDT-TTF)$_2$Cu(NCS)$_2$}
\def\khg{$\alpha$-(BEDT-TTF)$_2$KHg(SCN)$_4$}
\def\nh4{$\alpha$-(BEDT-TTF)$_2$NH$_4$Hg(SCN)$_4$}
\def\tlhg{$\alpha$-(BEDT-TTF)$_2$TlHg(SCN)$_4$}

\def\my{$\beta''$-(BEDT-TTF)$_4$[(H$_3$O)$M$(C$_2$O$_4$)$_3$]$\cdot Y$}
\def\pyr{$\beta''$-(BEDT-TTF)$_4$[(H$_3$O)$M$(C$_2$O$_4$)$_3$]$\cdot$C$_5$H$_5$N}

\def\cl2{(BEDT-TTF)$_3$Cl$_2 \cdot$ 2H$_2$O}

\def\Cl{$\kappa$-(BEDT-TTF)$_2$Cu[N(CN)$_2$]Cl}

\newcommand{\gtsim}{\mbox{{\raisebox{-0.4ex}{$\stackrel{>}{{\scriptstyle\sim}}
$}}}}
\newcommand{\ltsim}{\mbox{{\raisebox{-0.4ex}{$\stackrel{<}{{\scriptstyle\sim}}
$}}}}

\usepackage{graphicx}
\usepackage{dcolumn}
\usepackage{amsmath}
\usepackage{longtable}
\begin{document}

\title{Fermi-surface topology and the effects of intrinsic disorder
in a class of charge-transfer salts
containing magnetic ions,
$\beta''$-(BEDT-TTF)$_4$[(H$_3$O)$M$(C$_2$O$_4$)$_3$]$Y$}

\author{A.~I.~Coldea and A.~F. Bangura}
\affiliation{Clarendon Laboratory, University of Oxford, Parks
Road, Oxford OX1~3PU, UK}
\author{J. Singleton}
\affiliation{National High Magnetic Field Laboratory,
Los Alamos National Laboratory, TA-35, MS-E536, Los Alamos,
NM~87545 USA}
\author{A. Ardavan}
\affiliation{Clarendon Laboratory, University of Oxford, Parks
Road, Oxford OX1 3PU, UK}
\author{A. Akutsu-Sato\cite{japan}, H. Akutsu\cite{japan2}, S. S. Turner and P. Day}
\affiliation{Davy-Faraday Research Laboratory, The Royal
Institution, 21 Albemarle Street, London, W1S 4BS, UK}

\begin{abstract}
We report high-field magnetotransport measurements on
$\beta''$-(BEDT-TTF)$_4$[(H$_3$O)$M$(C$_2$O$_4$)$_3$]$Y$, where
$M=$ Ga, Cr and Fe and $Y=$  C$_5$H$_5$N. We observe similar
Shubnikov-de Haas oscillations in all compounds, attributable to
four quasi-two-dimensional Fermi-surface pockets, the largest of
which corresponds to a cross-sectional area $\approx 8.5\%$ of the
Brillouin zone. The cross-sectional areas of the pockets are in
agreement with the expectations for a compensated semimetal, and
the corresponding effective masses are $\sim m_{\rm e}$, rather
small compared to those of other BEDT-TTF salts. Apart from the
case of the smallest Fermi-surface pocket, varying the $M$ ion
seems to have little effect on the overall Fermi-surface topology
or on the effective masses. Despite the fact that all samples show
quantum oscillations at low temperatures, indicative of Fermi
liquid behavior, the sample- and temperature-dependence of the
interlayer resistivity suggest that these systems are
intrinsically inhomogeneous. It is thought that intrinsic tendency
to disorder in the anions and/or the ethylene groups of the
BEDT-TTF molecules leads to the coexistence of insulating and
metallic states at low temperatures; comparison with other
charge-transfer salts suggests that this might be a rather general
phenomenon. A notional phase diagram is given for the general
family of \my ~salts, which exhibits several marked similarities
with that of the $\kappa-$(BEDT-TTF)$_2$X superconductors.
\end{abstract}

\today

\maketitle
\section{Introduction}
Superconducting charge-transfer salts of the molecule BEDT-TTF have attracted
considerable experimental and theoretical interest because of their complex
pressure-temperature ($P$,$T$) phase diagrams, some of which are superficially
similar to those of the ``high-$T_{\rm c}$'' cuprate
superconductors~\cite{mck97,kanoda,jandc}. For example, the superconducting
phase in the $\kappa$-(BEDT-TTF)$_2X$ salts is in close proximity to an
antiferromagnetic insulator~\cite{french,nmr2,ito} and/or Mott
insulator~\cite{Limelette2003}; it is also surrounded by other unusual
states~\cite{french,nmr2}, including what has been termed a ``bad
metal''~\cite{Limelette2003}. Recent magnetisation~\cite{Sasaki2002}, thermal
expansion~\cite{Muller2002} and resistivity~\cite{Singleton2003} experiments
suggest that this ``bad metal'' may in fact represent the {\it coexistence} of
Fermi-liquid-like and insulating phases. The presence of both metallic and
insulating states at low temperatures is probably related to progressive
freezing-in of disorder associated with the terminal ethylene-groups of
BEDT-TTF (which can adopt either a ``staggered'' or ``eclipsed'' configuration)
and/or with the anions,
$X$~\cite{Muller2002,Akutsu2000,Sato,Tanatar2002,maksi}. As yet there is no
strong theoretical concurrence on the mechanism for superconductivity in the
BEDT-TTF salts~\cite{jandc,brandow}, with electron-electron interactions, spin
fluctuations~\cite{kuroki}, charge fluctuations~\cite{mckcf} and
electron-phonon interactions~\cite{vare} under consideration. It is therefore
unclear as to whether the mixed insulating/metallic phase referred to above is
a prerequisite for or a hindrance to superconductivity. However, a recent paper
has pointed out the sensitivity of the superconductivity in BEDT-TTF salts to
non-magnetic impurities and disorder, suggesting that this is evidence for
d-wave superconductivity~\cite{powell}.

In order to address some of these issues we have studied a new family of
charge-transfer salts of the form \my, ~where $M$ is a magnetic [Cr$^{3+}$
($S=3/2$), Fe$^{3+}$ ($S=5/2$)] or non-magnetic [Ga$^{3+}$ ($S=0$)] ion and $Y$
is a solvent molecule such as C$_5$H$_5$N (pyridine), C$_6$H$_5$CN
(benzonitrile) or C$_6$H$_5$NO$_2$ (nitrobenzene). $Y$ essentially acts as a
template molecule, helping to stabilize the structure; its size and
electronegativity affect the unit cell volume, and the amount of disorder in
the
system~\cite{Kurmoo1995,Akutsu2002JACS,Turner1999,Bangura2003,Martin2001,Rashid2001}.
The unit-cell volume is also affected by changing the $M$ ion inside the
tris(oxalate) structure~\cite{Akutsu2002JACS,Turner1999,Martin2001,Rashid2001}.
Furthermore, a subsidiary motive for varying $M$ is to search for potential
role for magnetism in the mechanism for superconductivity~\cite{Kurmoo1995}. In
this context, the magnetic charge-transfer salt
$\lambda$-(BETS)$_2$FeCl$_4$~\cite{uji,balicas} has been found to exhibit a
field-induced superconducting state in fields $\gtsim ~17$~T. Whilst these data
appear to be explicable by the Jaccarino-Peter compensation
effect~\cite{uji,jp,Uji2002}, others have suggested that the Fe ions play some
role in the superconducting state~\cite{balicas,cepas}.

\begin{figure}[htbp]
\centering
\includegraphics[width=7.5cm]{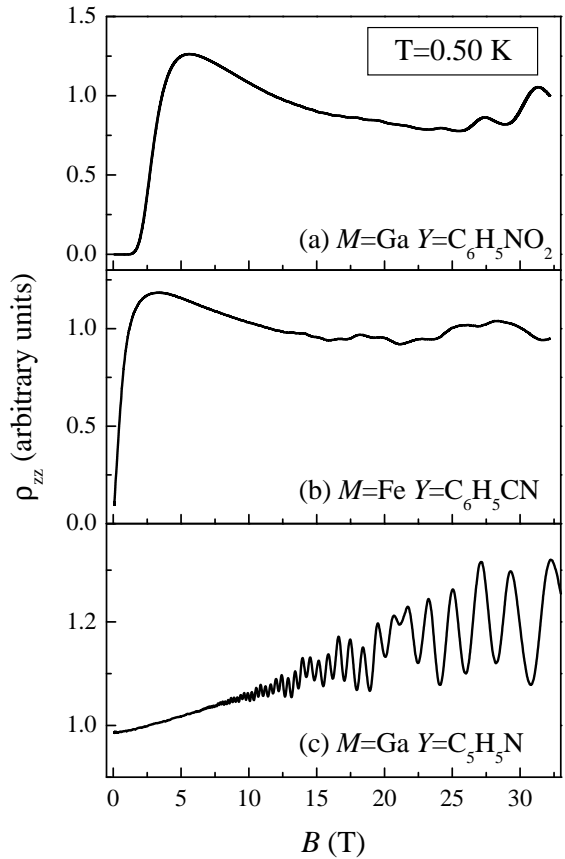}
\caption{Magnetic-field dependence of the interplane resistivity, $\rho_{zz}$
for \my ~samples containing different $Y$ molecules
and transition metal ions $M$.
Data are shifted vertically for clarity.
Salts with (a) $Y$=C$_6$H$_5$NO$_2$ or (b) $Y=$ C$_6$H$_5$CN
typically exhibit superconductivity, negative magnetoresistance and
a simple set of Shubnikov-de Haas oscillations.
By contrast, the (c) $Y=$ C$_5$N$_5$N (pyridine)
salt shows no superconductivity, positive
magnetoresistance and a complex series of Shubnikov-de Haas oscillations;
this is entirely typical of the salts containing pyridine.
}
\label{fig1}
\end{figure}
Although there are many detailed differences between individual samples, the
\my ~salts show two distinct classes of low-temperature behavior, as summarised
in Fig.~\ref{fig1}, which shows the interlayer magnetoresistivity $\rho_{zz}$
(see Section~\ref{experiment}) of three samples at a temperature $T=0.50$~K.
Salts with $Y=$ C$_6$H$_5$CN (benzonitrile) or C$_6$H$_5$NO$_2$ (nitrobenzene)
are superconductors~\cite{Kurmoo1995,Bangura2003}.
At temperatures above the
superconducting-to-normal transition,
they tend to exhibit negative magnetoresistance, on which is
superimposed one or two series of Shubnikov-de Haas oscillations of relatively
low frequency~\cite{Bangura2003}. On the other hand, salts with $Y=$
C$_5$H$_5$N are not superconducting; they exhibit positive magnetoresistance,
and display a complex mixture of higher-frequency Shubnikov-de Haas
oscillations. In this paper we shall concentrate on the $Y=$ C$_5$N$_5$N salts,
deriving their Fermi-surface parameters and quasiparticle scattering rates; the
superconductors with $Y=$ C$_6$H$_5$CN or C$_6$H$_5$NO$_2$ are described in
detail in another paper~\cite{Bangura2003}. However, in deriving a general
phase diagram (Section~\ref{discussion}) we shall discuss the latter
superconducting materials in general terms alongside the $Y=$ C$_5$H$_5$N
salts.

This paper is organised as follows. Experimental details
are given in Section~\ref{experiment};
relevant structural details and the behavior of the
\pyr ~samples on cooling from room to cryogenic
temperatures are given in Section~\ref{structure},
which also outlines the mechanisms which introduce disorder.
Magnetoresistance data are analysed in Section~\ref{magnetoresistance};
the Shubnikov-de Haas oscillations suggest that there are four Fermi-surface
pockets, the areas of which obey the additive relationship expected for
a compensated semimetal.
The results are discussed in Section~\ref{discussion};
this Section contains a notional phase diagram
for the \my ~salts which
shows the influence of unit cell size and disorder,
and which is compared with an equivalent phase diagram for the
$\kappa-$(BEDT-TTF)$_2$X salts.
A summary is given in Section~\ref{summary}.

\section{Experimental details}
\label{experiment} The \my ~samples were grown using
electrocrystallisation techniques as described
elsewhere~\cite{Kurmoo1995,Akutsu2002JACS,Turner1999}; they are
generally $\sim 1 \times 1 \times 0.2$~mm$^3$ hexagonal platelets or needles.
 It is possible to deduce the
upper and lower faces that are parallel to the highly-conducting
quasi-two-dimensional planes by visual inspection. Electrical
contacts were made to these surfaces by using graphite paint to
attach $12~\mu$m platinum wires. The interlayer
(magneto)resistance $R_{zz}$ $\propto \rho_{zz}$ (Ref.~\onlinecite{jandc})
was measured using standard four-terminal ac techniques. This
involves driving the current and measuring the voltage between
pairs of contacts on the upper and lower surfaces~\cite{jandc}.
Magnetoresistance experiments were carried out in quasistatic
fields provided by a superconductive magnet in Oxford and a 33~T
Bitter coil at NHMFL Tallahassee. The crystals were mounted in a
$^3$He cryostat which allowed rotation to all possible
orientations in magnetic field; sample orientation is defined by
the angle $\theta$ between the direction of the magnetic field and
the normal to the quasi-two dimensional planes and the azimuthal
angle $\phi$. Sample currents between 1 and 25~$\mu$A were used at
typical frequencies 18-300~Hz. Although around 20 crystals have
been studied, in this paper we shall focus on two or three typical samples of
each salt; samples are distinguished by the
consistent use of a label ({\it e.g.} $M=$~Cr, Sample~A).

\begin{table}
\caption{Lattice parameters of
\my
~salts ($C2/c$ symmetry group) measured
around 120~K. }
\begin{tabular}{lccccccc}
\hline \hline
  $M$/$Y$  & $a$(\AA) & $b$(\AA) & $c$(\AA) & $\beta$ & $V$(\AA$^3$) & $T$(K) & Ref.\\
  \hline
 Ga/C$_6$H$_5$NO$_2$ & 10.278 & 19.873 & 35.043 & 93.423 & 7145.2 & 100 & \cite{Akutsu2002JACS}\\
 Cr/C$_6$H$_5$NO$_2$ & 10.283 & 19.917 & 34.939 & 93.299 & 7144.4 & 150 &\cite{Rashid2001}\\
 Fe/C$_6$H$_5$NO$_2$ & 10.273 & 19.949 & 35.030 & 92.969 & 7169.6 & 120 &\cite{Rashid2001}\\
 \hline
 Cr/C$_6$H$_5$CN & 10.240 & 19.965 & 34.905 & 93.69 & 7121.6 & 120 & \cite{Martin2001}\\
 Fe/C$_6$H$_5$CN & 10.232 & 20.043 & 34.972 & 93.25 & 7157 & 120 & \cite{Kurmoo1995}\\
 \hline
 Ga/C$_5$H$_5$N & 10.258 & 19.701 & 34.951 & 93.366 & 7051.9 & 120 & \cite{Akutsu2002JACS}\\
 Fe/C$_5$H$_5$N  & 10.267 & 19.845 & 34.907 & 93.223 & 7101.0 & 150 &\cite{Turner1999}\\
 \hline
 \hline
\end{tabular}
\label{latpar}
\end{table}

\section{Structural considerations and disorder in the low-temperature
phase}
\label{structure}
\subsection{Structure and bandfilling}
\begin{figure}[htbp]
\centering
\includegraphics[width=8cm]{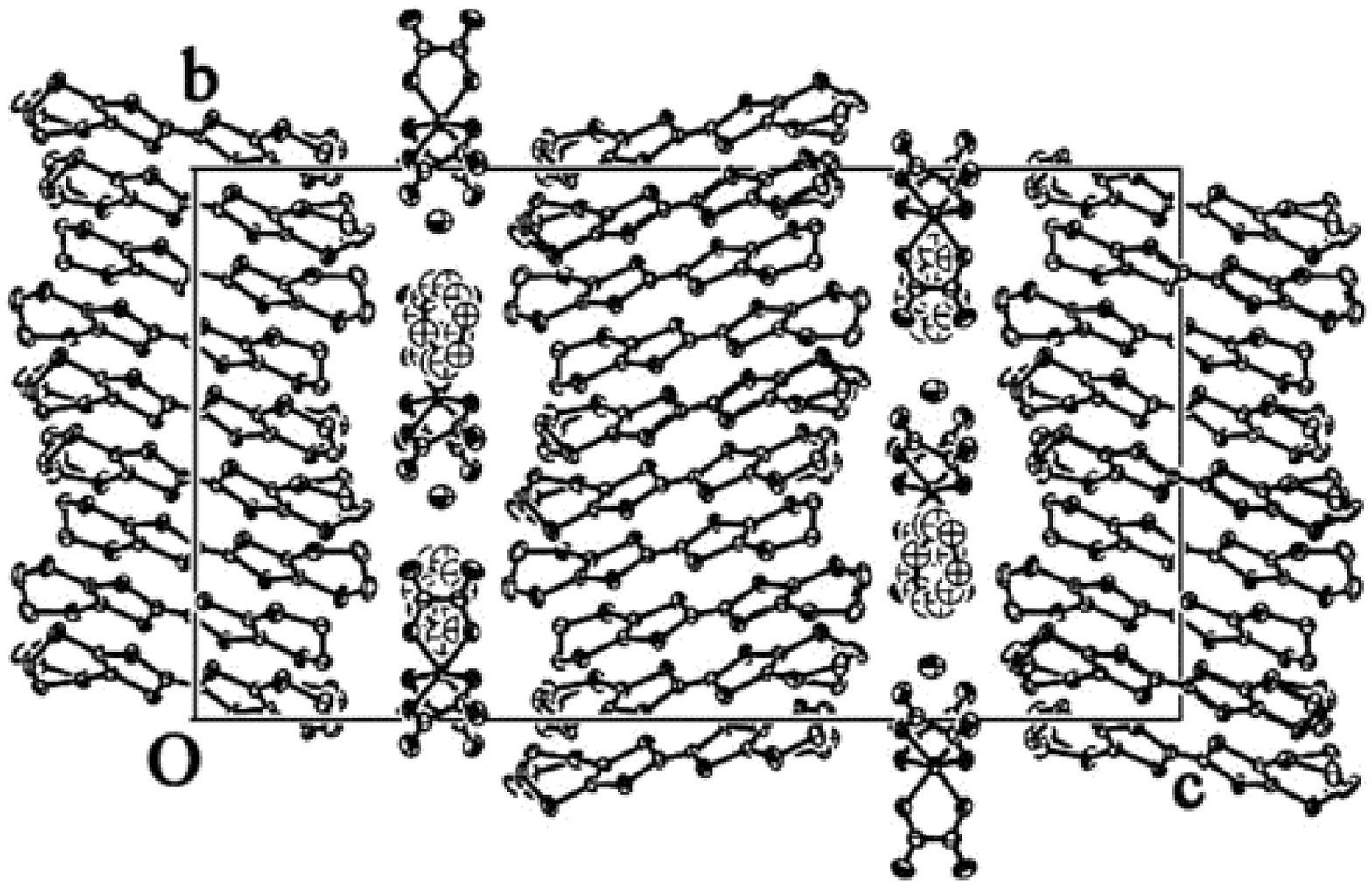}
\caption{Monoclinic crystal structure of
\pyr
~projected along the
$a$ axis \cite{Akutsu2002JACS}.}
\label{pyrstruct}
\end{figure}

Figure~\ref{pyrstruct} shows a projection of the crystal structure
along the $a$ axis of the \pyr ~salts, and Table~\ref{latpar} gives the lattice
parameters (around 120~K) for all compounds studied in this paper and in
Ref.~\onlinecite{Bangura2003}. The structure consists of alternating
BEDT-TTF and anion layers, the latter containing the metal
tris(oxalate) [$M$(C$_2$O$_4$)$_3$]$^{3-}$, the ion H$_3$O$^+$ and
the solvent molecule, $Y$. The molecules in the anion layer lie in
a ``honeycomb'' arrangement with alternate H$_3$O$^+$ and metal
oxalates giving an approximately hexagonal network of cavities in
which the solvent molecule $Y$ lies. The solvent molecule
 helps to stabilize the structure; the plane
 of phenyl ring makes
an angle of $\approx 32-36^{\circ}$ to the plane
of the oxalate layer~\cite{Martin2001,Turner1999,Rashid2001}. The
metal ion $M$ is octahedrally co-ordinated to the oxalate ligands;
the oxygen atoms on the oxalates are weakly bonded to the hydrogen
atoms on the terminal ethylene groups of the BEDT-TTF molecules,
acting to pull these together. The BEDT-TTF molecules
adopt the $\beta''$ packing arrangement in the $ab$ planes, in
which they form roughly orthogonal stacks. The crystallographic
structure of our compounds is monoclinic (see Table~\ref{latpar})
with the ($ab$) conducting planes at a distance of $d=c/2$ from
each other, as shown in Figure~\ref{pyrstruct}~\cite{Turner1999}.

By far the shortest S-S distances are within the cation planes,
leading to a predominantly two-dimensional
bandstructure~\cite{Kurmoo1995,Canadell2002}. Each BEDT-TTF molecule is
expected to donate half an electron, leaving two holes per unit
cell. Band structure calculations based on the room temperature
crystallographic data suggest these salts should be compensated
semimetals, with a Fermi surface consisting of
quasi-two-dimensional electron and hole pockets of approximately
equal area
~\cite{Canadell2002}. Although
BEDT-TTF salts and their relatives are frequently compensated
semimetals~\cite{jandc}, the electron-like Fermi-surface component
is often a pair of open sheets;  a closed electron pocket is
relatively unusual, but it was found in
$\beta''$-(BEDO-TTF)$_2$ReO$_4 \cdot $ H$_2$O~\cite{Kahlich1994}.

The interlayer transfer integrals will be less straightforward
to calculate in the \my ~salts; the planes of the
BEDT-TTF molecules in adjacent layers
(as well as those of the anion layers) are
twisted with respect to each other by $62 \pm 2^\circ$,
an unusual feature in BEDT-TTF salts \cite{Kurmoo1995,Martin2001}.

\subsection{Disorder mechanisms}
The \my ~salts are prone to structural disorder
primarily because the terminal ethylene
groups ($-$CH$_2$CH$_2 -$) of the BEDT-TTF
molecules are able to adopt
different configurations (twisted/staggered
or eclipsed)
depending on how they interact
with the anion layer~\cite{Turner1999,Akutsu2002JACS}.
Moreover, since C$_5$H$_5$N is smaller than the
other templating $Y$ molecules, it does not
fill the whole of the hexagonal cavity.
Changing the solvent molecule from
$Y=$ C$_6$H$_5$NO$_2$ to $Y=$ C$_5$H$_5$N
induces additional structural freedom,
leading to disorder in around
one quarter of the terminal ethylene
groups~\cite{Turner1999,Akutsu2002JACS}.
As a result, the ethylene groups
are the dominant cause of
both static and dynamic disorder at high
temperatures, and static disorder below
90~K, the temperature around which the
two different configurations are ``frozen in''~\cite{Akutsu2002JACS},
as found in the $\kappa-$phase salts~\cite{Muller2002,Akutsu2000,Sato}.

The C$_5$H$_5$N molecule can also introduce
disorder by adopting two different
orientations in the anion layer.
By contrast, the other solvents, $Y$= C$_6$H$_5$NO$_2$
and $Y$= C$_6$H$_5$CN, lock into
one ordered configuration~\cite{Akutsu2002JACS}.

Having discussed the various mechanisms for disorder,
we shall now examine how disorder is manifested in the
resistivity of the samples.
\begin{figure}[htbp]
\centering
\includegraphics[width=8cm]{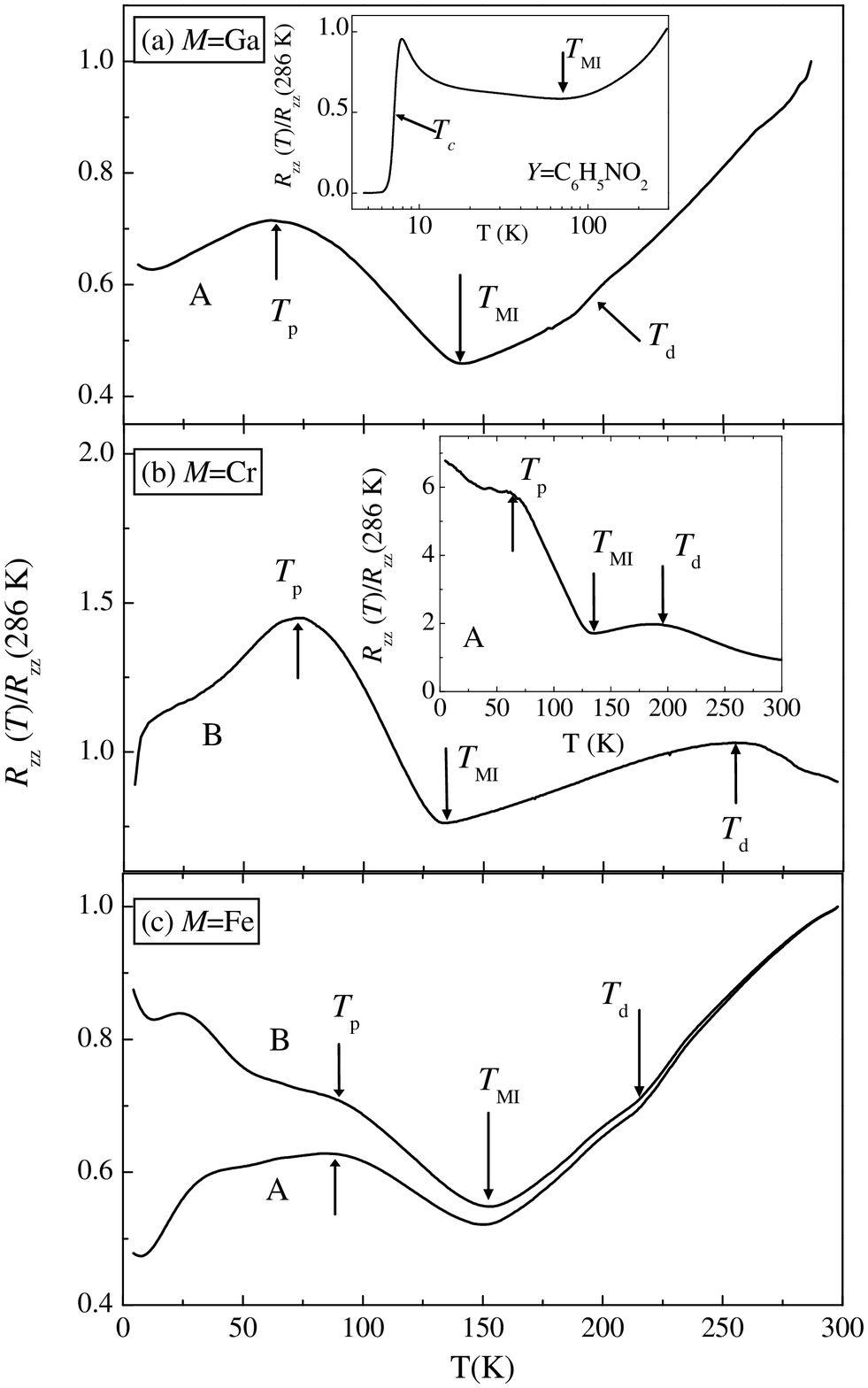}
\caption{Temperature ($T$) dependence of the
normalized interplane resistance
$R_{zz}$(T)/$R_{zz}$(286~K) in
zero magnetic field for different samples of
\pyr ~with (a)~$M=$ Ga (sample A) (the inset
shows \my ~with $M=$ Ga, $Y=$ C$_6$H$_5$NO$_2$
for comparison), (b)~$M=$ Cr sample B (the inset
shows $M=$ Cr sample A) and
(c)~$M=$ Fe (samples A and B).
The arrows indicate the temperatures
described in the text. }
\label{pyrRvsT}
\end{figure}

\subsection{The temperature dependence of the resistivity}

The temperature dependence of the normalized interplane resistance,
$R_{zz}(T)/R_{zz}$(286~K),
for five typical \pyr ~samples is
shown in Figure~\ref{pyrRvsT};
for comparison, equivalent data for $M=$ Ga, $Y=$ C$_6$H$_5$NO$_2$
are displayed in the inset.
Whilst many of the features in the data are
quite sample- or cooling-rate-dependent, all of the
samples ($M=$ Ga, Cr, Fe) are consistent in displaying a transition from
metallic-type behavior (positive ${\rm d}R_{zz}/{\rm d}T$)
to insulating-type behavior (negative ${\rm d}R_{zz}/{\rm d}T$)
at $T_{\rm MI} \approx 150$~K. Values of
$T_{\rm MI}$ are listed in Table~\ref{FSpar}.

The minimum in resistance at $T_{\rm MI}$ may represent the onset
of a possible form of density-wave state. Quasi-two-dimensional
conductors in which the Fermi surface is completely gapped by a
density wave exhibit a resistivity that rises by several orders of
magnitude as the temperature falls, as found for \cl2
~\cite{lub}). By contrast, the resistance of the \pyr ~salts
(shown in Figure~\ref{pyrRvsT}) only increases by a factor $\sim 1.5-3$. The
latter behavior is similar to that of quasi-two-dimensional
conductors in which a density wave only partially nests the Fermi
surface, leaving behind residual Fermi-surface pockets; examples
include the Mo bronzes~\cite{goddard} and \khg ~\cite{sasaki}. In
such cases, the conductivity is a convolution of a metallic
component, typically varying as a power-law in temperature (due to
the unnested portions of the Fermi surface) and an insulating
component with an activated temperature dependence (due to the
energy gap of the density-wave state)~\cite{goddard}.  The exact
form of the resistivity depends on which component
dominates.
An alternative scenario that could potentially lead to
similar resistivity behavior is the segregation of the sample into
insulating and metallic domains \cite{Dobrosavljevic2002},
as also proposed for the
$\kappa-$(BEDT-TTF)$_2$X salts (see Ref.~\onlinecite{Singleton2003} and
references therein). In
Section~\ref{magnetoresistance} we shall see that the
Fermi-surface topology is more complicated than that predicted by
the bandstructure calculations, which may be additional evidence
that the transition at $T_{\rm MI}$ is associated with the
formation of a density-wave.

All of the $Y$=C$_5$H$_5$N crystals
also consistently exhibit a feature
at a lower temperature, $T_{\rm p}\approx 60-80$~K
(shown in Fig.~\ref{pyrRvsT}).
However, depending on the sample,
this is manifested either as
a change from insulating- to metallic-type behavior
($M=$ Ga, all samples, $M=$ Cr, sample B,
$M=$ Fe sample A), or as merely
a shoulder on a resistivity that
continues to increase with decreasing temperature
($M=$ Cr, sample A, $M=$ Fe, sample B).
Such a feature is also indicative of
a number of contributions to the
conductivity acting in parallel.
for example, it is possible to reproduce
the behavior of $M=$ Cr sample B between 60~K
and $T_{\rm MI}$ using a
resistor network model that combines metallic (resistivity
$\propto T^n$, with $n \sim 1-2$)
and thermally-activated
components $\propto \exp(E_{\rm A}/k_{\rm B}T)$
(see also Ref.~\onlinecite{activationCr}).
Although the exact values obtained depend on the
details of the resistor network model used,
the values of $E_{\rm A}$ obtained from fitting
data between $T_{\rm p}$ and $T_{\rm MI}$
showed a consistent increase from
$M=$ Fe ($E_{\rm A} \approx 170-220$~K)
through $M=$ Ga ($E_{\rm A} \approx 300$~K)
to $M=$ Cr ($E_{\rm A} \approx 400-500$~K),
{\it i.e.} the activation energy
$E_{\rm A}$ increases with decreasing
unit cell volume (see Table~\ref{latpar}).

The features discussed thus far do not seem to depend
on sample cooling rate.
By contrast, in all five $M=$Cr samples studied, there is
an additional peak in the resistivity
at $T_{\rm d}\approx 200-270$~K,
the appearance and temperature of which
both depend on the sample cooling rate.
By contrast, samples with $M=$Ga, Fe
only exhibit a small inflection at $T_{\rm d}$.
At the lowest temperatures, $R_{zz}(T)/R_{zz}(286$~K)
values ranging from around 0.5 ($M=$ Fe, sample A)
to 7 ($M=$ Cr, sample A) are obtained (Fig.~\ref{pyrRvsT});
the actual value reached seems more dependent on the sample
batch rather than the identity of the $M$ ion
({\it e.g.} compare $M=$ Cr samples A and B).
This points to a prominent role for disorder
in determining the low-temperature resistive behavior
of the \pyr ~salts.

As $T$ tends to zero, the resistivity of
$M=$ Cr sample B drops quite sharply,
although zero resistance is never attained.
A similar drop in resistance for
$M=$Ga below 2~K, which was destroyed
by an applied field of 0.16~T, was previously
reported as evidence for superconductivity~\cite{Akutsu2002JACS}.
However, none of the $M=$ Ga samples studied in the
present work exhibited such a feature.
This is possibly related to the recent observation
that superconductivity in the BEDT-TTF salts is
very sensitive to disorder and non-magnetic impurities~\cite{powell}.

On the other hand, a robust superconducting state is stabilized
below $T_{\rm c}=7$~K for $M=$ Ga and $Y=$ C$_6$H$_5$NO$_2$
(as shown in the inset of Figure~\ref{pyrRvsT}(a)) and
for $M$= Fe and $Y$= C$_6$H$_5$CN (Figure~\ref{fig1}(b)
and Ref.~\onlinecite{Kurmoo1995}).
For completeness, note that
both of the latter superconducting salts
show a single metal-insulator transition (see inset of Figure~\ref{pyrRvsT}(a))
similar to that observed at $T_{\rm MI}$ in
the $Y=$ C$_5$H$_5$N salts.
However, for the superconducting
salts $T_{\rm MI}$ seems somewhat sample dependent;
 values ranging from $T_{\rm MI}=$68~K~\cite{Bangura2003}
to $T_{\rm MI} \approx 160-180$~K~\cite{Akutsu2002JACS} have
been reported for the
$M$=Ga, $Y$=C$_6$H$_5$NO$_2$ salt.

To summarise this section, the resistivities
of the \pyr ~salts exhibit a complex temperature
and sample dependence (Figure~\ref{pyrRvsT}).
The minimum in $R_{zz}$ at $T_{\rm MI}$ is an intrinsic feature of
all samples, and, by analogy with resistivity data
from other quasi-two-dimensional systems, probably indicates
the onset of a density-wave state.
The form of the resistivity at temperatures just below this
(including the peak at $T_{\rm p}$)
suggests metallic and thermally-activated
contributions to the conductivity
acting in parallel. At lower temperatures,
the behavior of the samples is much more divergent,
with $R_{zz}(T)/R_{zz}(286)$ values spread between $0.5$
and 7 indicating an additional
thermally-activated process (or processes) which is (are)
probably dependent on the degree of disorder within the samples.
By contrast, the temperature-dependent resistivity
is rather simpler for the \my ~salts
with $Y$= C$_6$H$_5$NO$_2$ and $Y$= C$_6$H$_5$CN$_2$.
The difference may be attributable to the
higher degree of structural disorder possible in
the $Y=$ C$_5$H$_5$N salts, resulting from the
less constrained ethylene groups and greater rotational
freedom of the $Y$ molecule~\cite{Akutsu2002JACS}.
Similar electronic properties determined by the
disordered anions (that lock into two different
configurations) were found for
$\beta''$-(BEDT-TTF)$_2$SF$_2$CHFCF$_2$SO$_3$ \cite{Schlueter2001b},
for which resistivity
shows a metal-insulating
transition near 190~K,
compared with the superconducting compound,
$\beta''$-(BEDT-TTF)$_2$SF$_2$CF$_2$CF$_2$SO$_3$ ($T_c=5.4$~K),
which has ordered anions  \cite{Jones2000}.

\begin{figure*}[htbp]
\centering
\includegraphics[height=7.9cm]{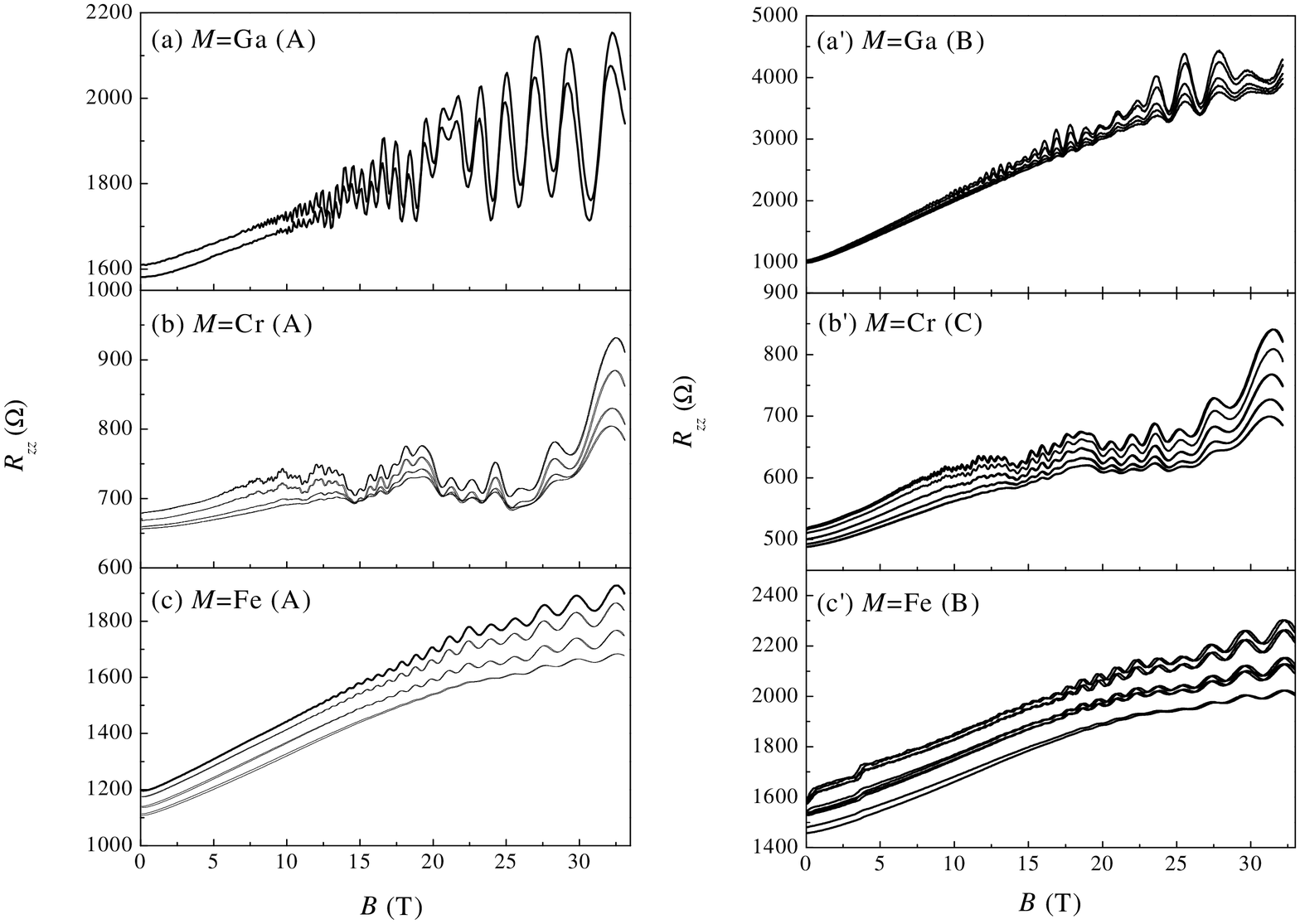}
\includegraphics[height=8.1cm]{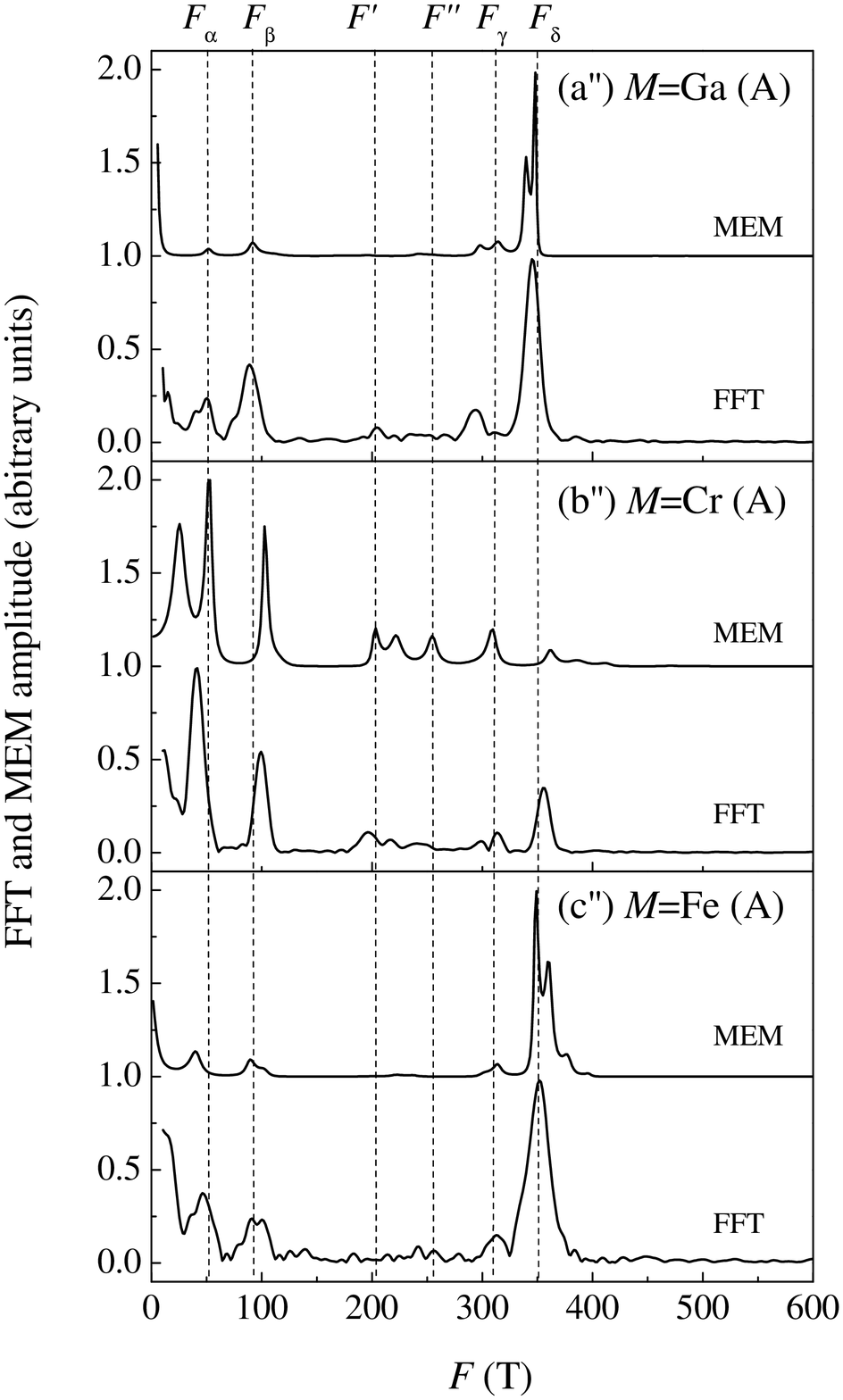}
\caption{Magnetic field dependence of
the interplane resistance, $R_{zz}$,
for \pyr ~samples
$M=$ Ga (samples A (a) and B (a')),
$M=$ Cr (samples A (b) and C (b'))
and $M=$~Fe (samples A (c) and B (c')),
recorded between $T=0.5-4.2$~K.
The right panel [(a"), (b") and (c")]
corresponds to the  maximum entropy method (top solid lines)
and Fast Fourier transform spectra (bottom solid lines)
of the oscillatory component of the resistance,
[$(R_{zz}-R_{\rm bg})/R_{\rm bg}$,
where $R_{\rm bg}$ is a polynomial fit] over the field range 7-32~T;
the transforms correspond to the data from
left panel at $T=0.5$~K.
The dashed vertical lines indicate the approximate
positions of the frequencies discussed in the text.}
\label{pyrRvsB}
\end{figure*}
\section{Low-temperature magnetoresistance}
\label{magnetoresistance}
\subsection{Shubnikov-de Haas frequencies and Fermi-surface pockets}
Figure~\ref{pyrRvsB} shows the field dependence
of $R_{zz}$ for several samples of \pyr
~with $M=$ Ga, Cr and Fe measured at
several temperatures between 0.50~K and 4.2~K.
All samples exhibit Shubnikov-de Haas oscillations
superimposed on a positive
background magnetoresistance.
Several frequencies are visible in varying proportions.
For example, the dominant series of oscillations for $M=$ Cr
is of relatively low frequency, whereas the dominant
oscillations for $M=$ Ga, Fe are of a higher frequency.
The amplitude of the oscillations varies
slowly with temperature,
suggesting the corresponding effective masses are
not very large~\cite{Shoenberg}.

No clear signature of
superconductivity was observed either in the field, angle or
temperature dependence of $R_{\rm zz}$ when $Y=$ C$_5$H$_5$N, in
contrast to the situation in \my ~salts
having different solvents [as shown in Fig.~\ref{fig1} and inset
of Figure~\ref{pyrRvsT}(a)].

In order to analyse the Shubnikov-de Haas oscillations,
we define the oscillatory fraction of the magnetoresistance,
\begin{equation}
\frac{\Delta R_{zz}}{R_{\rm bg}}= \frac{R_{zz}-R_{\rm bg}}{R_{\rm bg}}.
\end{equation}
Here $R_{\rm bg}$ is the slowly-varying background
magnetoresistance approximated by a polynomial in $B$.
As long as $\Delta R_{zz}/R_{\rm bg} \ll 1$,
$\Delta R_{zz}/R_{\rm bg} \approx -\Delta \sigma_{zz}/\sigma_{\rm bg}$,
where the $\sigma$ are equivalent terms in the conductivity~\cite{Shoenberg,review}
($\Delta \sigma_{zz}/\sigma_{\rm bg}$ is the quantity
dealt with in the Lifshitz-Kosevich (LK) treatment of Shubnikov-de Haas
oscillations~\cite{Shoenberg} used
to extract effective masses and
the scattering time of the quasiparticles).
The $\Delta R_{zz}/R_{\rm bg}$ values were processed
using both the maximum entropy method (MEM)
(filter size $=200$)~\cite{Sigfusson}
and the Fast Fourier transform (FFT) usually over the field range
$7-32$~T.
The two methods give similar
representations of the frequencies present, as shown in the
right panel of Fig.~\ref{pyrRvsB}.

\begin{figure}[htbp]
\centering
\includegraphics[width=6.6cm]{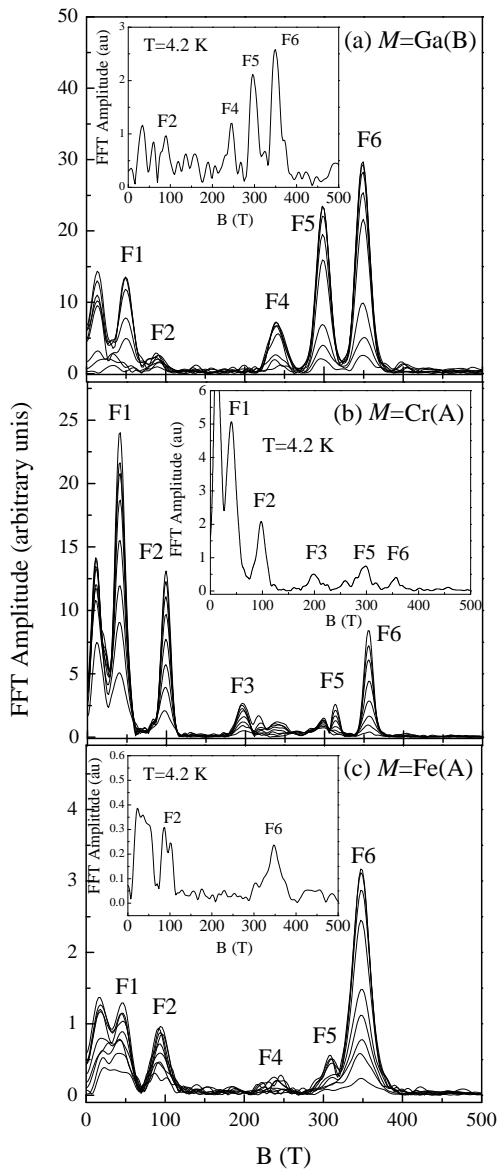}
\caption{Fast Fourier transforms of Shubnikov-de Haas oscillations for
samples of \pyr. ~(a)~$M=$ Ga (B),  (b)~$M=$ Cr (A) and (c)~$M=$
Fe recorded between $T=0.5$~K and $T=4.2$~K.} \label{pyrFT}
\end{figure}

We identify four frequencies
which occur consistently in all of the transforms
over the complete temperature range (see Figures~\ref{pyrRvsB}
and \ref{pyrFT}), and are similar
in all
\pyr ~samples with $M=$ Ga, Cr and Fe.
These frequencies are $F_{\alpha} \approx 38-50$~T,
$F_{\beta} \approx 86-98$~T,
$F_{\gamma} \approx 293-308$~T and $F_{\delta} \approx 345-353$~T;
the ranges cover the values observed in the different samples
(see Table~\ref{FSpar}).
In addition, two other peaks, with frequencies
$F' \approx 190-206$~T and $F'' \approx 236-248$~T,
were observed less consistently in the transforms.
The peak at very low frequencies ($\ltsim 20$~T)
is an artifact of the subtraction of the
background magnetoresistance;
its position and amplitude depends on whether
$R_{\rm bg}$ was approximated by a second or fourth-order
polynomial in $B$.
In some cases, the peak at frequency $F_{\rm \alpha}$
is superimposed on the flank of
this feature, making a precise determination
of the frequency difficult.

Because of their dependence on temperature (see below),
magnetic-field orientation (all
frequencies vary as $1/\cos \theta$, where $\theta$
is the angle between the magnetic field and
the normal to the quasi-two-dimensional planes~\cite{Coldea2003})
and their consistent appearance in the transforms, we
attribute the Shubnikov-de Haas frequencies
$F_{\alpha}$, $F_{\beta}$, $F_{\gamma}$ and $F_{\delta}$
to the extremal orbits about
four independent quasi-two-dimensional Fermi-surface pockets,
which we label $\alpha,~\beta,~\gamma$ and $\delta$.
As for the other peaks, we note that
as $F' \approx 2F_{\beta}$, it is likely to be a
second harmonic of the oscillations due to the $\beta$
pocket.

The peak seen occasionally in the transforms at
a frequency $F''$ seems likely to result from
frequency-mixing effects and
it can be constructed using a variety of recipes
(for example, $F'' \approx F_{\alpha}+2F_{\beta}$,
$F'' \approx F_{\gamma}-F_{\alpha}$,
$F''\approx F_{\delta}-F_{\beta}$).
Such frequency-mixing effects in quasi-two-dimensional
metals are often attributable to the chemical potential
becoming pinned to relatively sharp Landau levels over restricted
regions of magnetic field (the so-called ``chemical potential
oscillation effect'' (CPOE))~\cite{eva,neilbd}, which,
in some cases, very complex mixed
harmonics are generated~\cite{oldsingleton}.
Another possibility which can generate a difference
frequency is the Stark Quantum Interference
effect~\cite{review};
this represents ``interference'' of two semiclassical
Fermi-surface orbits between which tunnelling can occur.
However, the oscillations due to the Stark Quantum
Interference effect are usually characterised
by an apparent very light effective mass; that is,
their amplitude varies more slowly with temperature than that
of the oscillations due to the two ``parent'' orbits~\cite{kartsovnik}.
The fact that, when present, the oscillations at $F''$
are suppressed much more rapidly with increasing temperature than
any of the possible parent frequencies suggests that
CPOE is the more likely explanation~\cite{magint}.

At this point, it is worth recalling that the bandstructure
calculations predict only two Fermi-surface pockets, of equal
area~\cite{Canadell2002}, whereas the experimental data
suggest four pockets. There are several potential reasons for this
difference. Firstly, whilst extended-H\"{u}ckel calculations often
give a reasonable qualitative description of the Fermi surfaces of
many BEDT-TTF salts~\cite{jandc}, the $\beta''$-phases have proved
problematic; slight differences in input parameters seem to result
in wildly-differing predicted topologies (see, for example, the case
of $\beta''-$(BEDT-TTF)$_2$AuBr$_2$~\cite{house}). Secondly, the
bandstructure calculations are based on
structural measurements carried out at relatively high
temperatures~\cite{Akutsu2002JACS}; contraction of the lattice
could result in changes in the relative sizes of the various
transfer integrals, leading to shifts in the bands with respect to
the chemical potential.
Finally, the presence of a series of
pockets could be a consequence of a Fermi surface reconstruction
determined by a possible charge-density wave
at $T_{\rm MI}$ of the \pyr ~salts.
 Similar Fermi-surface reconstructions have been
suggested for other $\beta''$ salts, including
$\beta''$-(BEDT-TTF)$_2$AuBr$_2$ (where a plethora of small Fermi
surface pockets results)~\cite{house},
$\beta''$-(BEDO-TTF)$_2$ReO$_4 \cdot $ H$_2$O~\cite{Kahlich1994}
and \betasf, ~where it appears that the Fermi-surface nesting is
more efficient~\cite{msn}.

In spite of the larger number of Fermi-surface
pockets observed experimentally, there
are some similarities with the calculated Fermi surface.
First, the largest experimental pocket, $\delta$,
is of a similar cross-sectional
area ($\approx 8.5$\% of the Brillouin cross-section)
to the calculated pockets
($8.1$\% of the Brillouin-zone
cross-section)~\cite{Canadell2002}.
Secondly, as noted above, the \my ~salts
are expected to be quasi-two-dimensional compensated semimetals
in which the cross-sectional areas of the hole
Fermi-surface pockets should sum to the same value as the total
cross-sectional area of the electron Fermi-surface pockets.
We note that $F_{\alpha}+F_{\delta} \approx F_{\beta} + F_{\gamma}$
to reasonable accuracy (Table~\ref{FSpar}).
This suggests that if $\alpha$ and $\delta$ are
electron (hole) pockets, then $\beta$ and $\gamma$
will be hole- (electron-) like.

Although the Shubnikov-de Haas oscillation frequencies
are generally similar for the three \pyr ~salts,
there are detail differences depending on the ion $M$.
For example, the $F_{\alpha}$ frequency
of the $M=$ Cr salts is consistently lower than
that of the $M=$ Ga and Fe compounds.
The appearance of the Shubnikov-de Haas oscillations
is also affected by the $M$ ion;
the highest frequency oscillations ($F_{\delta}$)
dominate the spectra of the compounds
with $M=$ Ga and Fe,
whereas that of the compounds with $M=$ Cr
is dominated by the low frequency, $F_{\alpha}$
(see Figures~\ref{pyrRvsB} and~\ref{pyrFT}).
This may be related to relatively
small differences in the scattering mechanisms,
rather than some intrinsic effect of the Cr$^{3+}$ ion.
Examples of similar effects were observed in magnetoresistance
data for the low-field, low-temperature phases
of \khg and \tlhg ~\cite{eva2}.
The relative amplitudes of the various
Shubnikov-de Haas oscillation series vary from
sample to sample, and batch to batch, with some
series being undetectable in what is presumed to
be the lower-quality samples, whilst
being relatively strong in other crystals
(see Sections 1 and 5 of
Ref.~\onlinecite{eva2} and references cited therein).
A second example is
$\beta''-$(BEDT-TTF)$_2$AuBr$_2$
for which comparison
of the magnetic-quantum oscillation data from
Refs.~\onlinecite{house,doporto,house10,house11}
shows that the relative amplitudes of the lower
and higher-frequency oscillation
series varies considerably from sample
to sample.

For completeness, we mention that the superconducting salts,
$Y=$ C$_6$H$_5$NO$_2$ with  $M=$ Ga and Cr
and $Y=$ C$_6$H$_5$CN with $M=$ Fe
show only two frequencies, with the low
frequency in the range $47-55$~T and
the high frequency in the range $190-238$~T~\cite{Bangura2003}.

\subsection{Effective masses and Dingle temperatures}
A two-dimensional Lifshitz-Kosevich formula~\cite{neilmodel}
has been used to extract the effective masses $m^*$ of the
various Fermi-surface pockets, where possible.
The Fourier amplitude of each series of
quantum oscillations is given by
\begin{equation}
A_{\rm 2D} \propto R_TR_{\rm D}R_S,
\label{LK}
\end{equation}
where $R_T=\frac{X}{\sinh(X)}$ is the temperature damping term,
$R_{\rm D}=\exp\left(-X \frac{T_{\rm D}}{T}\right)$
is the Dingle term ($T_{\rm D}$ is the Dingle temperature)
due to the broadening of the Landau levels
caused by internal inhomogeneities
and $X=14.694 \frac{T}{B} \frac{m^*}{m_{\rm e}}$.

The spin-splitting term
$R_S=\left |  \cos\left(\frac{\pi}{2}
\frac{g^* m^*/m_{\rm e}}{\cos(\theta)}\right)\right |$,
where $g^ *$ is the effective g-factor,
but is not considered
here and it will be the
subject of a future publication~\cite{Coldea2003}.

\begin{figure}[htbp]
\centering
\includegraphics[width=6cm]{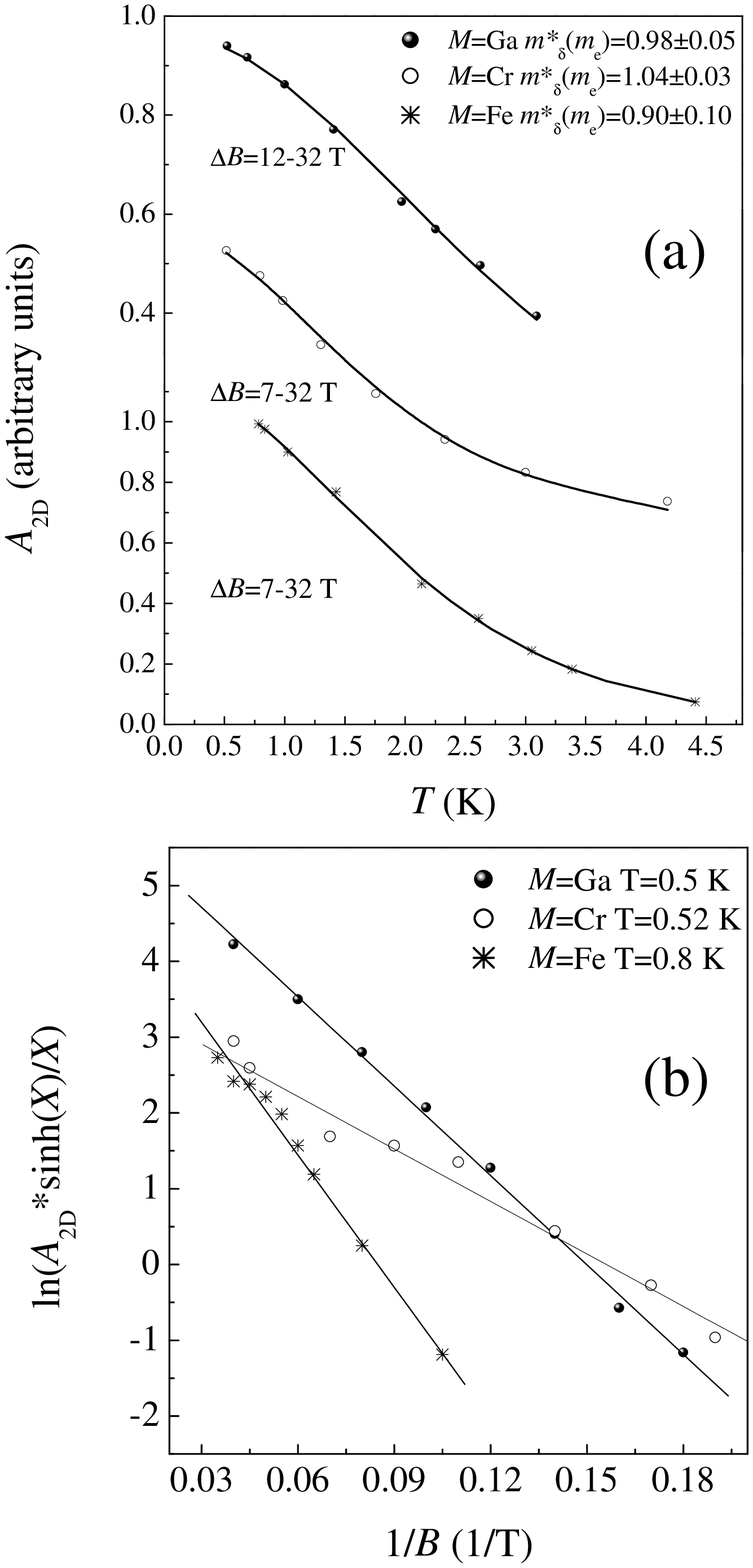}
\caption{(a) Temperature dependence of the Fourier amplitude
of the $F_{\delta}$ frequency in
\pyr ~for different $M$.
The solid line is a fit to the data (points) using the
$R_T$ term of Eq.~\ref{LK} (solid line).
The field window was $\Delta B=7-32$~T for the samples with $M=$ Cr (sample A)
and $M=$ Fe (sample A)
and $\Delta B=12-32$~T for the sample with $M=$ Ga (sample A).
Data for different $M$
are offset for clarity.
(b) The corresponding Dingle plots (ln[$A_{\rm 2D} \sinh(X)/X$]
versus $1/B$,
where $X=14.695~m^*T/B$)
for the $F_{\delta}$ frequency at $T \approx 0.5$~K.
The solid line is a linear fit to the data (points).
The field windows overlapped by less than $\approx 30$\%.}
\label{pyrm}
\end{figure}

The Fourier amplitudes obtained over a field
window $7-32$~T were fitted to the $R_T$ term of Eqn.~\ref{LK},
using around 8 different temperatures covering the
range $0.5-4.2$~K
(for consistency, a
polynomial of the same order was used to subtract the background
magnetoresistance for each sample).
Figure~\ref{pyrm} shows
typical amplitudes and corresponding fits
for the $F_{\delta}$ series.
All of the $m^*$ values obtained
for the different Fermi-surface pockets
are listed in Table~\ref{FSpar}.

To the limit of experimental error
the effective masses for the $\gamma$ and $\delta$ pockets
of the three salts are
close to the free-electron mass, $m_{\rm e}$.
Whilst such values are light compared to the
typical masses observed in
\betasf~\cite{msn} or the $\kappa-$ and $\alpha-$phase
BEDT-TTF salts~\cite{review}, they are
not without precedent in charge-transfer
salts~\cite{Kahlich1994,lub}.
The effective masses of the $\alpha-$pocket
are somewhat smaller for the $M=$ Cr and Fe salts
($m^* \approx m_{\rm e}/2$); however, in the
case of the $M=$ Ga salt, the $\alpha$ effective mass
seems rather larger.
Apart from this, there is yet no evidence
that the magnetic moment on the 3d ions $M=$ Cr and Fe
has any effect on the effective masses.
This is in contrast to the study
on $\kappa$-(BETS)$_2$FeCl$_4$,
where it was proposed that spin fluctuation effects
enhanced the effective mass~\cite{neilbets}.
On the other hand, the effective mass in
$\lambda$-(BETS)$_2$Fe$_x$Ga$_{1-x}$ Cl$_4$
is not very much affected by the presence of
the magnetic ions but it is much larger
than that in our compounds ($\approx 4 m_e$ ~\cite{Uji2002}).

\begin{table*}[htbp]
\caption{Parameters associated
with the bandstructure of
\pyr ~for magnetic fields perpendicular to the highly-conducting
quasi-two-dimensional planes.
Values for several samples with different $M$ are given.
The $F$ are Shubnikov-de Haas frequencies, the subscripts
$\alpha$ {\it etc.} identifying the associated Fermi-surface
pocket; the $m^*$ are corresponding effective masses. $T_{\rm D \delta}$
is the Dingle temperature for $F_{\delta}$, and $T_{\rm MI}$
is the metal-insulator transition temperature identified in
Fig.~\ref{pyrRvsT}. }
\centering
\begin{tabular}{lcccccccc}
\hline
\hline
Parameters  &$M=$Ga(A)&  $M=$Ga(B)& $M=$Ga(C) &$M=$Cr(A)& $ M=$Cr(B) &$M=$Cr (L) &$M=$Fe(A) &$M=$Fe(B) \\
\hline
$F_{\alpha}$~(T) & 48  &  50   &  49  & 39   &  38  &  40   & 45    &45  \\
$F_{\beta}$~(T)  & 89  &  85   &  92  & 95   &  95  &  98   & 94    & 92  \\
$F'$ (T)         & 205 &  --   &  --  & 190  &  190 &  195  & --    & --  \\
$F''$ (T)        & 247 &  240  & 235  & --   &  --  &  243  & 243   & -- \\
$F_{\gamma}$ (T) & 292 &  296  & 297  & 296  &  286 &  305  & 307   & 305  \\
$F_{\delta}$ (T) & 344 &  345  & 346  & 344  &  343 &  357  & 346   & 344 \\
\hline
$m^*_{\alpha}$($m_{\rm e}$) & -- & $1.9\pm 0.3$ &$1.3\pm0.2$ &$0.56\pm0.05$  & $0.54\pm0.05$  & $0.5\pm0.1$  & $0.8\pm0.1$ & $0.6\pm0.1$\\
$m^*_{\beta}$($m_{\rm e}$) &$0.56 \pm 0.05$  & $0.51 \pm 0.05$  &$0.62 \pm 0.05$  &$0.63\pm0.05$  & $0.62\pm0.05$  & --           &$0.68\pm0.05$ & $0.76\pm0.05$\\
$m^*_{\gamma}$($m_{\rm e}$) &$0.7 \pm 0.1$   & $1.01 \pm 0.05$  &$1.09 \pm 0.05$  & --   & --             &--            & --           & --\\
$m^*_{\delta}$($m_{\rm e}$) &$0.98 \pm 0.05$  & $0.95 \pm 0.05$  &$0.93 \pm 0.05$  &$1.04\pm0.05$  & $0.98\pm0.05$  &$0.9\pm0.1$   & $0.9\pm0.1$  & $1.1\pm0.1$\\
\hline
$T_{\rm D \delta}$ (K) & $2.7\pm0.1$ & $2.3\pm0.2$    &   $1.7\pm0.2$ &  $1.8\pm0.1$ & $1.4\pm0.2$& $2.5\pm0.5$ & $4\pm0.5$ & $4.2\pm0.1$\\
\hline
$T_{\rm MI}$(K) & $138 \pm 2$   & -- & -  &  $142\pm 1$  & -- &  $120\pm20$   & $150\pm 2$ &  $153\pm 2$\\
\hline
\hline
\label{FSpar}
\end{tabular}
\end{table*}

A further insight into the properties of our samples
is given by the Dingle temperature, $T_{\rm D}$, which
can be used to parameterise the scattering rate~\cite{powell,Shoenberg},
the spatial potential fluctuations or a combination of the
two~\cite{Singleton2003,powell}.
The $T_{\rm D}$ values for the $\delta$ pocket are
listed  in Table~\ref{FSpar}; typical fits are shown in
Fig.~\ref{pyrm}(b).
Note that $T_{\rm D}$
is consistently larger for the compounds with
$M=$Fe ($T_{\rm D} \approx 4$~K, corresponding to
a scattering time of $\tau \approx 0.3$~ps)
and is smaller for the salts with
$M=$Cr ($T_{\rm D} \approx 1.5$~K,
corresponding to $\tau \approx 0.8$~ps).
This difference is visible even in the raw data,
with fewer oscillations being visible for the $M=$Fe salt.
As both compounds
contain magnetic ions, some form of magnetic scattering
(such as spin-disorder scattering~\cite{kusters})
may be excluded as the reason for these differences;
it is more likely to be related
to the degree of nonmagnetic disorder present,
determined by the anions and the solvent.

Interestingly, there is no apparent correlation between
the values of $R_{zz}(T)/R_{zz}(286$~K) (see Figure~\ref{pyrRvsT})
and the Dingle temperatures for each sample (Table~\ref{FSpar}).
For example, the sample with the largest $R_{zz}(T)/R_{zz}(286$~K)($\approx 7$)
($M=$ Cr, sample A) has a $T_{\rm D}$ which is a factor 2.2
smaller than that of the sample with the smallest
$R_{zz}(T)/R_{zz}(286$~K)($\approx 0.5$) ($M=$ Fe, sample A).
The Dingle temperatures extracted from Shubnikov-de Haas oscillations
suggest that
$M=$ Cr sample A is of higher quality, whereas $M=$ Fe sample
A has the lower resistivity ratio.
This strongly suggests that the samples are not
of a uniform single phase at the lowest temperatures
but their overall properties probably
represent a mixture of metallic and insulating domains.
Within this mixture, the metallic domains may well be of
quite high quality, as evidenced by the observation
of Shubnikov-de Haas oscillations with a reasonably small Dingle temperature.

Further support for such an idea is given by
comparing the values of $R_{zz}(10~$K$)/R_{zz}(286$~K)$\sim 0.5-7$
seen in Fig.~\ref{pyrRvsT}
with $R_{zz}(10~$K$)/R_{zz}(286$~K)$\sim 0.001$ obtained
for the unambiguously metallic
salt $\beta-$(BEDT-TTF)$_2$I$_3$~\cite{tokumoto}.
This great disparity is an indication that
a large fraction of the
quasiparticles in the \pyr ~salts
that are mobile at room
temperature do not contribute to the bulk conductivity
at low temperatures.
This loss of charge-carriers is presumably be related to the
suggested density-wave transition at $T_{\rm MI}$
(which perhaps gaps part of the Fermi surface) and
to the subsequent ``freezing out'' of further quasiparticles
(suggested by the negative ${\rm d} R_{zz}/{\rm d}T$
values seen for several of the samples as shown in Figure~\ref{pyrRvsT})
caused by disorder at lower temperatures.

\section{Discussion: proposed phase diagram}
\label{discussion}

The previous sections have described the
transport properties of
 \pyr ~salts
exhibiting a metal-insulator transition
at $T_{\rm MI}$ (probably associated with a density-wave
state) and Shubnikov-de Haas oscillations
at lower temperatures, indicative of a
reasonably good metal. However, depending on
the sample batch, the overall
resistivity can be much greater than that at
room temperature; the most likely explanation is that
the sample consists of a mixture of metallic
and insulating domains. The tendency for a particular
region of the sample to remain metallic or become
insulating may be linked to particular
configurations of the anion and/or ethylene groups
possible in the \pyr ~salts (see the discussion
of intrinsic disorder in Section~\ref{structure}).

These findings are summarised in
Figure~\ref{phd}(a),
which shows a notional phase diagram for
all of the \my ~salts as a function of ``chemical pressure''
($=-\Delta V/V$),
{\it i.e.} the fractional difference in unit-cell
volume of a particular salt from that of the
$M=$ Fe, $Y=$ C$_6$H$_5$NO$_2$ compound, which has
the largest unit cell. For comparison,
Figure~\ref{phd}(7) shows an analogous
diagram for the $\kappa-$(BEDT-TTF)$_2$X salts
(after Ref.~\onlinecite{Singleton2003}
based on Refs.~\onlinecite{french,Sasaki2002,Muller2002}).
There are a number of quite striking similarities between the
two phase diagrams.
\begin{enumerate}
\item
The superconductivity is suppressed by the
reduction in volume of the unit cell.
The suppression of superconductivity
is accompanied by the increasingly ``metallic''
character of both families
of materials; in the \my salts, this is evidenced
by the increase in the number and size of Fermi-surface pockets
observed (two for $Y=$ C$_6$H$_5$NO$_2$, C$_6$H$_5$CN,
four, generally larger ones for $Y=$ C$_5$H$_5$N);
in the $\kappa$-(BEDT-TTF)$_2$X salts this shows up as
an increase in the Shubnikov-de Haas oscillation frequencies
and the low-frequency optical conductivity~\cite{jandc}.
This trend is confirmed by hydrostatic pressure
measurements of the
superconducting
$\beta"$-{BEDT-TTF}$_4$[(H$_3$O)Ga(C$_2$O$_4$)$_3$]$\cdot Y$
salts, which showed that the superconductivity is
destroyed and the number of Fermi-surface pockets
increased by increasing pressure~\cite{Volodia2002}.
\item
As the chemical pressure increases,
the suggested density-wave transitions
(which occurs at $T_{\rm MI}$ in the \my ~salts and at $T^*$ in the
$\kappa-$(BEDT-TTF)$_2$X compounds) increases in
temperature, at least initially.
\begin{figure}[htbp]
\centering
\includegraphics[width=7.5cm]{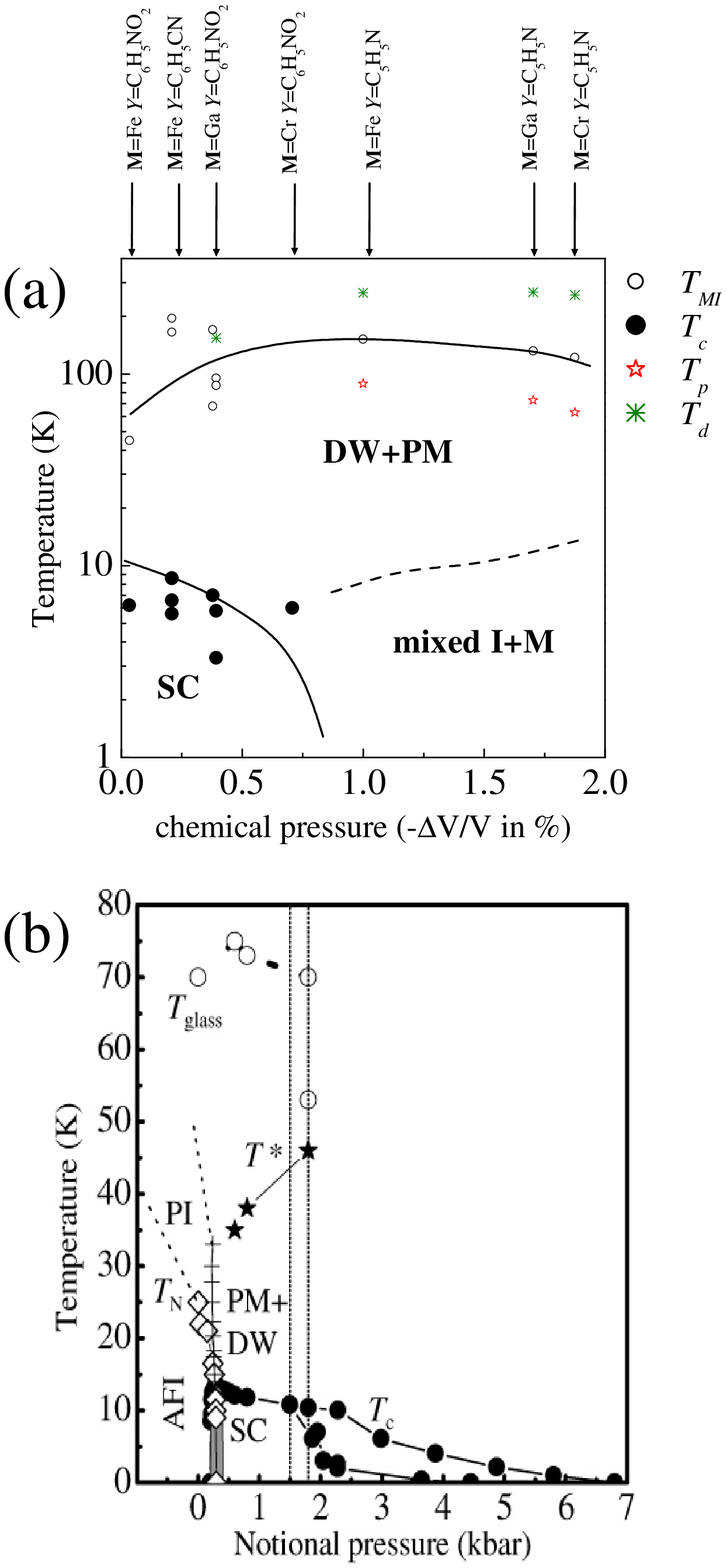}
\caption{(a) Notional phase diagram
of the \my ~salts, using data from the current paper and
from Ref.~[23].
~Solid circles correspond to the superconducting
critical temperature, $T_{\rm c}$,
open circles represent the metal-insulator transition, $T_{\rm MI}$.
The other temperatures, $T_{\rm p}$ and
$T_{\rm d}$, are described in the text.
Different phases are: SC=superconducting,
DW-density wave, M-metallic, PM-paramagnetic metallic
and I-insulating phase.
The solid and dashed lines are guides to the eye.
(b) Phase diagram of $\kappa$-(BEDT-TTF)$_2$X
including boundaries suggested by recent data
``Notional pressure'' combines chemical pressure
caused by changing anion X
and applied hydrostatic pressure;
``$0$'' is ambient pressure for \Cl; ~the vertical
lines are the ambient pressure positions
of deuterated (left) and  undeuterated (right) \cuscn
(after
Ref. [9] based on Refs.~[4,8,10]).}
\label{phd}
\end{figure}
In this context, recall that
the characteristic activation
energy $E_{\rm A}$ (Section~\ref{structure})
also increases with chemical pressure.

\item
In both families, the superconducting state
is surrounded by regions in which metallic and
insulating behavior coexist ``PM + DW'' phase (in the
$\kappa$-phase salts after Ref.~\cite{Sasaki2002}).
\end{enumerate}

The \my ~salts seem to emphasise
two emerging themes in the physics of organic
(super)conductors.
Firstly, there have been a number of recent instances in which
very high sample resistivity and phenomena indicative of
quasiparticle localisation ({\it e.g.} Anderson localisation~\cite{goddard2})
and/or disorder~\cite{musfeldt}
coexist with effects normally associated with
``good metals'', such as Shubnikov-de Haas oscillations~\cite{storr} or even
superconductivity~\cite{goddard2}.
Secondly, there is experimental evidence that the precursor
to superconductivity may involve the coexistence of
metallic and density-wave-like states
(for example \betasf~\cite{msn} or
It has been suggested in the
case of the $\kappa-$phase salts~\cite{Singleton2003}
that these phases exist in
distinct ``domains'' or regions of the sample,
the behavior of a particular domain being determined
by local structural arrangements.

Finally, recent theoretical work has emphasised
the role of disorder in the suppression
of superconductivity in (BEDT-TTF) salts.
Often a measure of this disorder is derived
from Shubnikov-de Haas-oscillation or
cyclotron-resonance data~\cite{powell}.
The resistivity data indicate that disorder
makes some regions
of the samples prone to localisation
and these regions contribute
 little to the low-temperature conductivity.
Other regions remain metallic and exhibit Shubnikov-de Haas
oscillations, indicative of reasonably long scattering times
and mean-free paths $\sim 300~$\AA~
and hence low disorder.
Thus it is important to
emphasise that
Shubnikov-de Haas
and cyclotron resonance data are only informative
about the disorder in the metallic regions of
a sample (see also Ref.~\onlinecite{Singleton2003}).

\section{Summary}
\label{summary}
In conclusion we have studied
the Fermi-surface topology of
\my, ~ (with $M=$ Ga, Cr, Fe and $Y=$C$_5$H$5$N).
All of the studied salts exhibit similar
Shubnikov-de Haas-oscillation spectra,
which we attribute to four quasi-two-dimensional
Fermi-surface pockets.
The cross-sectional areas of the pockets are
in agreement with the expectations for a compensated
semimetal, and the corresponding
effective masses are $\sim m_{\rm e}$,
rather small compared to those of
other BEDT-TTF salts.
Apart from the case of the
smallest Fermi-surface pocket, varying the $M$ ion
seems to have little
effect on the overall Fermi-surface topology
or on the effective masses.

Despite the fact that all samples show quantum
oscillations at low temperatures,
indicative of Fermi liquid behavior,
the sample- and temperature-dependence
of the interlayer resistivity
lead us to suggest that these systems are
intrinsically inhomogeneous.
It is thought that intrinsic tendency
to disorder in the anions and/or the ethylene
groups of the BEDT-TTF molecules
leads to phase separation of the samples
into insulating and metallic states.

Based on the data in this paper,
and those from Ref.~\onlinecite{Bangura2003},
we have constructed a notional phase diagram
for the \my ~salts which exhibits
several similarities with that
of the $\kappa-$(BEDT-TTF)$_2$X superconductors,
and which could
have larger implications
for other charge-transfer salts.

\begin{acknowledgments}
This work is supported by EPSRC (UK), the Royal Society
(UK), INTAS (project number 01-2212) and by the US
Department of Energy (DoE) under grant LDRD-DR 20030084.
Work at the National High Magnetic Field Laboratory is performed
under the auspices of the National Science Foundation, the
State of Florida and DoE.
We are grateful to Prof. E. Canadell for sending us
the results of his bandstructure
calculations.
We thank Drs N. Harrison and J. Lashley for fruitful discussions,
and Drs A.-K. Klehe and V. Laukhin for
useful comments and access to their recent high-pressure data.
\end{acknowledgments}


\begin{thebibliography}{10}
\bibitem[$\dag$]{japan} Present address:
Department of Materials Science,
Graduate School and Faculty of Science,
Himeji Institute of Technology, Hyogo 678-1297, Japan
\bibitem[$\ddag$]{japan2} Present address: Department of Organic and Polymeric
Materials, Tokyo Institute of Technology, Tokyo 152-8552, Japan


\bibitem{mck97}
R.~H. McKenzie, Science {\bf 278},  820  (1997).
\bibitem{kanoda}
K.~Kanoda, Physica C {\bf 282-287}, 299 (1997)
\bibitem{jandc}
J. Singleton and C.H. Mielke, Contemp. Phys. {\bf 43}, 63 (2002).
\bibitem{french}
S. Lefebvre, P. Wzietek, S. Brown,
C. Bourbonnais, D. Jerome, C. Meziere,
M. Fourmigue and P Batail,
Phys. Rev. Lett. {\bf 85}, 5420 (2000).
\bibitem{nmr2}
K. Miyagawa, A. Kawamoto and K. Kanoda,
Phys. Rev. Lett. {\bf 89}, 017003 (2002).
\bibitem{ito}
H. Ito, G. Saito and T. Ishiguro,
J. Phys. Chem. Solids {\bf 62}, 109 (2001).
\bibitem{Limelette2003}
P. Limelette, P. Wzietek, S. Florens,
A. Georges, T.A. Costi, C. Pasquier, D. Jerome,
C. Meziere and P. Batail,
cond-mat/0301478  (2003).
\bibitem{Sasaki2002}
T. Sasaki, N. Yoneyama, A. Matsuyama, and
N. Kobayashi, Phys. Rev. B {\bf 65}, 060505  (2002).
\bibitem{Singleton2003}
J. Singleton, C.~H. Mielke, W. Hayes,
and J.~A. Schlueter, J. Phys.:
Condens. Matter {\bf 15},  L203  (2003).
\bibitem{Muller2002}
J. M\"{u}ller, M. Lang, F. Steglich,
J.A. Schlueter, A.M. Kini and T. Sasaki,
Phys. Rev. B {\bf 65},  144521  (2002).
\bibitem{Akutsu2000}
H. Akutsu, K. Saito, and M. Sorai,
Phys. Rev. B {\bf 61},  4346  (2000).
\bibitem{Sato}
A. Sato, H. Akutsu, K. Saito and M. Sorai,
Synth. Metals {\bf 120}, 1035 (2001).
\bibitem{Tanatar2002}
M.~A. Tanatar, T. Ishiguro, S. Kagoshima,
N.D. Kushch, and E.B. Yagubskii,
Phys. Rev. B {\bf 65},  064516  (2002).
\bibitem{maksi}
M. Maksimuk, K. Yakushi, H. Taniguchi, K. Kanoda
and A. Kawamoto, cond-mat/0305680 (2003).
\bibitem{brandow}
B.H. Brandow, Phil. Mag. (in press) (2003).
\bibitem{kuroki}
K. Kuroki, T. Kimura, R. Arita, Y. Tanaka and
Y. Matsuda, Phys. Rev. B {\bf 65}, 100516 (2002);
H. Kondo and T. Moriya, J. Phys.: Condens. Matter
{\bf 11}, L363 (1999).
\bibitem{mckcf}
J. Merino and R.H. McKenzie, Phys. Rev. Lett.
{\bf 87}, 237002 (2001); M. Calandra,
J. Merino and R.H. McKenzie,
Phys. Rev. B {\bf 66} 195102 (2002).
\bibitem{vare}
G. Varelogiannis, Phys. Rev. Lett.
{\bf 88}, 117005 (2002).
\bibitem{powell}
B.J. Powell and R.H. McKenzie,
cond-mat/0306457 (2003).
\bibitem{Kurmoo1995}
M. Kurmoo, A.W. Graham, P. Day, S.J. Coles,
M.B. Hursthouse, J. Caulfield, J. Singleton,
F.L. Pratt, W. Hayes, L. Ducasse
and P.J. Guionneau,
J. Am. Chem. Soc. {\bf 117},  12209  (1995).
\bibitem{Akutsu2002JACS}
H. Akutsu, A. Akutsu-Sato, S.S. Turner, D. Le Pevelen,
P. Day, V. Laukhin, A.-K. Klehe, J. Singleton,
D.A. Tocher, M.R. Probert and J.A.K. Howard,
J. Am. Chem. Soc. {\bf 124},  12430  (2002).
\bibitem{Turner1999}
S.~S. Turner, P. Day, K.M.A. Malik,
M.B. Hursthouse, S.J. Teat, E.J. Maclean,
L. Martin and S.A. French,
Inorg. Chem. {\bf 38},  3543  (1999).
\bibitem{Bangura2003}
A. Bangura {\it et~al.}, in preparation (2003);
A. F. Bangura, A. I. Coldea, J. Singleton,
A. Ardavan, A. K. Klehe, A. Akutsu-Sato, H. Akutsu,
S. S. Turner, P. Day, Synthic Metals, {\bf 1-3},  1313  (2003).
\bibitem{Martin2001}
L. Martin, S.S. Turner, P. Day, P. Guionneau,
J.A.K. Howard, K.M.A. Malik, M.A. Hursthouse,
M. Uruichi and M. Yakushi,
Inorg. Chem. {\bf 40},  1363  (2001).
\bibitem{Rashid2001}
S. Rashid, S.S. Turner,
P. Day, J.A.K. Howard, P. Guionneau,
E.J.L. McInnes, F.E. Mabbs, R.J.H. Clark,
S. Firth and T.J. Biggs,
J. Mater. Chem {\bf 11},  2095  (2001).
\bibitem{uji}
S. Uji, H. Shinagawa, T. Terashima,
T. Yakabe, T. Terai, M. Tokumoto, A.
Kobayashi, H. Tanaka and H. Kobayashi,
Nature {\bf 410},  908  (2001).
\bibitem{balicas}
L. Balicas, J.S. Brooks, K. Storr,
S. Uji, M. Tokumoto, H. Tanaka,
H. Kobayashi, A. Kobayashi,
V. Barzykin and L.P. Gorkov,
Phys. Rev. Lett. {\bf 87},  067002  (2001).
\bibitem{jp}
V. Jaccarino and M. Peter,
Phys. Rev. Lett. {\bf 9},  290  (1962).
\bibitem{Uji2002}
S. Uji, C. Terakura, T.Terashima, T. Yakabe,
Y. Terai, M. Tokumoto, A. Kobayashi,
F. Sakai, H. Tanaka, H. Kobayashi,
Phys. Rev. B {\bf 65}, 113101 (2002).
\bibitem{cepas}
O. Cepas, R.H. McKenzie and J. Merino, Phys.
Rev. B {\bf 65}, 100502 (2002).
\bibitem{Canadell2002}
T. G. Prokhorova, S. S. Khasanov, L. V. Zorina, L. I. Buravov,
V. A. Tkacheva, A. A. Basakakov, R. B. Morgunov, M. Gerer,
E. Canadell, R. P. Shibaeva and E. B. Yagubskii,
Adv. Funct. Mat., {\bf 13}, 403 (2003).
\bibitem{Kahlich1994}
S. Kahlich, D. Schweitzer, C. Rovira, J.A. Paradis,
M.H. Whangbo, I. Heinen, H.J. Keller, B. Nuber,
P. Belle, H. Brunner and R.P. Shibaeva,
Z. Phys. B {\bf 60},  3060  (1999).
\bibitem{lub}
W. Lubczynski, S.V. Demishev, J. Singleton,
J.M. Caulfield, L. du Croo de Jongh, C.J. Kepert,
S.J. Blundell, W. Hayes, M. Kurmoo and P. Day,
J. Phys.:Condens. Matter {\bf 8}, 6005  (1996).
\bibitem{goddard}
J. Dumas and C. Schlenker, Int. J. Mod. Phys.
B {\bf 7} 4045 (1993).
\bibitem{sasaki}
T. Sasaki, N. Toyota, M. Tokumoto,
N. Kinoshita and H. Anzai,
Solid State Commun. {\bf 75} 93 (1990).
\bibitem{Dobrosavljevic2002}
V. Dobrosavljevic, D. Tanaskovic and A. A. Pastor,
cond-mat/0206529 (2002).
\bibitem{activationCr}
 An additional
activated component must be introduced to reproduce the
data for $M=$ Cr sample A over the same temperature region.
\bibitem{tokumoto}
M. Tokumoto, I. Nishiyama, K. Murata,
H. Anzai, T. Ishiguro and G. Saito,
Physica B {\bf 143}, 372 (1986).
\bibitem{Schlueter2001b}
J. A. Schlueter, B. H. Ward, U. Geiser, H. H. Wang,
  A. M. Kini, J. Parakka, E. Morales, H.-J. Koo, M.-H. Whangbo,
  R. W. Winter, J. Mohtasham and G. L. Gard,
  J. Mater. Chem., {\bf 11},  2008 (2001).
\bibitem{Jones2000}
B. R. Jones, I. Olejniczak, J. Dong,
J. M. Pigos, Z. T. Zhu, A. D. Garlach,
J. L. Musfeldt, H.-J. Koo, M.-H. Whangbo,
J. A. Schlueter, B. H. Ward, E. Morales, A. M. Kini,
R. W. Winter, J. Mohtasham and G. L. Gard, Chem. Mater.,
{\bf 12}, 2490 (2000).
\bibitem{Shoenberg}
D. Shoenberg, {\em Magnetic Oscillations in Metals},
{\it Cambridge Monographs in Physics}
(Cambridge University Press, Cambridge, 1984).
\bibitem{Sigfusson}
T.~I. Sigfusson, K.~P. Emilsson,
and P. Mattocks, Phys. Rev. B {\bf 46}, 10446
(1992).
\bibitem{Coldea2003}
A.I. Coldea, in preparation (2003).
\bibitem{eva}
See {\it e.g.} E. Rzepniewski,
R.S. Edwards, J. Singleton, A. Ardavan
and Y. Maeno, J. Phys.: Condens.
Matter {\bf 14} 3759 (2002) and references
therein.
\bibitem{neilbd}
N. Harrison, J. Caulfield, J. Singleton,
P.H.P. Reinders, F. Herlach, W. Hayes,
M. Kurmoo and P. Day, J. Phys.: Condens.
Matter {\bf 8}, 5415 (1996).
\bibitem{oldsingleton}
J. Singleton, F. Nasir and R.J. Nicholas,
Proc. SPIE {\bf 659}, 99 (1986).
\bibitem{review}
J. Singleton, Rep. Prog. Phys.
{\bf 63}, 1111 (2000).
\bibitem{kartsovnik}
M.V. Kartsovnik, G. Yu. Logvenov,
T. Ishiguro, W. Biberacher, H. Anzai and
N.D. Kushch, Phys. Rev. Lett.
{\bf 77}, 2530 (1996).
\bibitem{magint}
Note that Shoenberg's mechanism for
frequency-mixing effects
due to the {\it oscillatory}
magnetic field within the sample~\cite{Shoenberg}
is not generally feasible in
BEDT-TTF salts, because the low
quasiparticle density results in
a rather small oscillatory
magnetisation~\cite{house}.
In this context, the $M$ ion is also unimportant;
our compounds do not show any long
range magnetic order down to 0.5~K \cite{ColdeaJMMM}.
\bibitem{house}
A.A. House, N. Harrison, S.J. Blundell,
I. Deckers, J. Singleton, F. Herlach, W. Hayes,
J.A.A.J. Perenboom, M. Kurmoo and P.Day,
Phys. Rev. B {\bf 53}, 9127 (1996).
\bibitem{ColdeaJMMM}
A.~I. Coldea {\it et~al.}, J. Mag. Mag. Mat.
submitted  (2003).
\bibitem{msn}
M.-S. Nam, A. Ardavan, J.A. Symington,
J. Singleton, N. Harrison, C.H. Mielke,
J.A. Schlueter, R.W. Winter and G.L. Gard,
Phys. Rev. Lett. {\bf 87}, 117001 (2001).
\bibitem{eva2}
N. Harrison, E. Rzepniewski, J. Singleton,
P.J. Gee, M.M. Honold, P. Day and
M. Kurmoo, J. Phys.: Condens. Matter
{\bf 11}, 7227 (1999).
\bibitem{doporto}
M. Doporto, J. Singleton, F.L. Pratt,
J. Caulfield, W. Hayes,
J.A.A.J. Perenboom, I. Deckers, G. Pitsi,
M. Kurmoo and P. Day, Phys. Rev. B
{\bf 49}, 3934 (1994).
\bibitem{house10}
S. Uji, H. Aoki, M. Tokumoto, A. Ugawa
and K. Yakushi, Physica B {\bf 194},
1307 (1994).
\bibitem{house11}
M. Tokumoto, A.G. Swanson, J.S. Brooks,
C.C. Agosta, S.T. Hannahs, N. Kinoshita,
H. Anzai, M. Tamura, H. Tajima,
N. Kuroda, A. Ugawa and K. Yakushi,
Physica B {\bf 184}, 508 (1993).
\bibitem{neilmodel}
N. Harrison, R. Bogaerts, P.H.P. Reinders,
J. Singleton, S.J. Blundell and F. Herlach,
Phys. Rev. B {\bf 54}, 9977 (1996);
N. Harrison, A. House, I. Deckers, J. Caulfield,
J. Singleton, F. Herlach, W. Hayes,
M. Kurmoo and P.Day, {\it ibid.} {\bf 52},
5584 (1995).
\bibitem{neilbets}
N. Harrison, C.H. Mielke, D.G. Rickel,
L.K. Montgomery, C. Gerst and J.D. Thompson,
Phys. Rev. B {\bf 57}, 8751 (1998).
\bibitem{kusters}
Y. Shapira, S. Foner and N.F. Oliveira,
Phys. Rev. B {\bf 10}, 4765 (1974).
\bibitem{Volodia2002}
V. Laukhin and A.-K. Klehe,
private communication (2002).
\bibitem{goddard2}
P. Goddard, S.W. Tozer, J. Singleton,
A. Ardavan, A. Abate and M. Kurmoo,
J. Phys.: Condens. Matter, {\bf 14}, 7345 (2002).
\bibitem{musfeldt}
I. Olejniczak, J.L. Musfeldt, G.C. Papavassiliou
and G.A. Mousdis, Phys. Rev. B {\bf 62}, 15634
(2000).
\bibitem{storr}
K. Storr, L. Balicas, J.S. Brooks, D. Graf
and G.C. Papavassiliou, Phys. Rev. B {\bf 64},
045107 (2001).

\end{thebibliography}
\end{document}